\documentstyle[sprocl]{article}

\def\ra{\rightarrow}

\def\bq{\begin{equation}}
\def\eq{\end{equation}}
\def\be{\begin{eqnarray}}
\def\ee{\end{eqnarray}}
\def\ben{\begin{enumerate}}\def\een{\end{enumerate}}
\def\d {\partial}

\def\np{Nucl. Phys.}\def\pl{Phys. Lett.}
\def\la{\langle}\def\ra{\rangle}

\def\roughly#1{\mathrel{\raise.3ex\hbox{$#1$\kern-.75em%
\lower1ex\hbox{$\sim$}}}}

\begin{document}

\title{THE CHESHIRE CAT REVISITED}

\author{V. VENTO}

\address{Departament de F\'{\i}sica Te\`orica - Institut de F\'{\i}sica 
Corpuscular, Universitat de Val\`encia- Consejo Superior de Investigaciones
Cient\'{\i}ficas, E-46100 Burjassot (Val\'encia), Spain\\E-mail:
vicente.vento@uv.es}


\maketitle\abstracts{The concept of effective field theory leads in a natural
way to a construction
principle for phenomenological sensible models known under the name of the
Cheshire Cat Principle.  We review its formulation in the chiral bag scenario 
and discuss its realization for the flavor singlet axial charge.\\
Quantum effects inside the chiral bag induce a color anomaly which 
requires a compensating surface term to prevent breakdown of color 
gauge invariance. The presence of this
surface term allows one to derive
in a gauge-invariant way a chiral-bag version of the Shore-Veneziano 
two-component formula for the flavor-singlet 
axial charge of the proton.  We show that one can obtain a striking 
Cheshire-Cat phenomenon with a negligibly small singlet axial charge.
}

\section{Introduction}
\indent\indent The possibility of  formulating  a physical theory by means of
equivalent field theories, defined in terms of different field variables, 
leads to a construction principle for phenomenological sensible models, which
will be the subject of our presentation. Named the Cheshire Cat Principle
($CCP$), by Nielsen and collaborators \cite{nielsen1,nielsen2}, for reasons 
which will become apparent shortly, it allows to confront experimental data 
with a restricted parameter set obtained by imposing theoretical consistency 
within the model.

$1 + 1$-dimensional fermionic theories are bosonizable \cite{coleman}
and thus an exact transformation relating the fermionic and bosonic fields may 
be defined \cite{mandelstam}, which allows the construction of the same
$S$-matrix from apparently different bosonic and fermionic theories. The
intricacies of describing fermions by bosonic fields results in the 
appearence of topologically non trivial scenarios in field theory. Since both 
descriptions lead to the same physics, it is a matter of simplicity and/or 
taste which language to use \cite{ferrando}. In terms of the $1+1$-dimensional
scenario, the 
Cheshire Cat Principle, as stated by Nielsen and collaborators, is exact and 
transparent, and therefore we shall discuss it in some detail in the next 
section.

In the real world, bosonization with a finite number of bosonic variables, 
is not exact. However the unproven theorem of Weinberg \cite{weinberg}
leading to  the formulation of low energy effective theories \cite{georgi}, 
supports the $CCP$. Moreover, the $CCP$ can be used as a consistency check for 
effective theories, allowing to fix parameters, which would 
otherwise have to be determined from the data, and therefore, the $CCP$,
increases 
the predictive 
power of the theories.

At sufficiently low energies or long distances Quantum Chromodynamics ($QCD$) 
can be described accurately by an effective field theory written in terms of
meson fields \cite{leutwyler,skyrme,witten}. In this regime, the color
fermionic description of the theory is extremely complex due to confinement. The
$CCP$ is operative when one defines $QCD$ in a two phase space-time scenario 
and matches appropriately two field theories. To be more specific, one
defines a hypertube in four dimensions, the bag, whose interior is governed by
conventional $QCD$ described in terms of quark and gluon fields. In the exterior
of the bag, one assumes a certain mesonic model satisfying the flavor 
symmetries of the theory and the basic postulates of field theory in accord with
Weinberg's theorem. The two descriptions are matched by using the appropriate
boundary conditions which implement the symmetries and confinement. The $CCP$
states that the physics should be approximately independent of the spatial size
of the bag.

The $CCP$ has been tested in many instances with notable success \cite{NRZ}.
There is one case, the flavor axial singlet charge ($FSAC$), were its 
implementation has not been succesful beyond doubt and therefore it has merit 
our attention \cite{PVRB,RV}.  From the phenomenological point of view the 
$FSAC$ is associated with the $\eta^\prime$ and therefore with the anomaly 
\cite{hooft}.
This observable is
relevant for what has been referred to as the {\it proton spin problem}. 
In the chiral bag model the formulation is very elaborate.
Confinement induces through quantum 
effects a color anomaly, which leads to a surface coupling of the $\eta^\prime$ 
with the gluon field. The latter induces a gauge non invariant Chern-Simons 
current, whose expectation value we need to calculate. We have shown \cite{RV},
and we
will review it in here, that the presence of the surface term generated by the
proper matching of the color anomaly with the surface gluon-$\eta^\prime$
coupling allows us to derive in a gauge invariant way a chiral-bag version of
the Shore-Veneziano two component formula for the $FSAC$  of the
proton \cite{shore}.

\section{Exact Cheshire Cat Principle in $1+1$ dimensions}
\indent\indent $1+1$-dimensional fermion theories are exactly bosonizable
\cite{coleman}. The $CCP$ arises when one implements bosonization in a bag
\footnote{We strongly advise to read the beautiful presentation of
ref.\cite{nielsen1}, where specially the graphics are extremely illuminating.
One may also turn for completeness and further bibliography to ref.\cite{NRZ}.}.

Let us consider a massless free single-flavored fermion $\psi$ confined to a
region
of {\it volume} V (inside) coupled on the surface $\partial V$ to a massless
free boson $\varphi$ living in a region of {\it volume} $\overline{V}$ (outside). 
Of
course in one space dimension, the {\it volume} is just a segment of a line but
we will use the symbol in analogy to higher dimensions. We will assume that the
action is invariant under global chiral rotations and parity. The action
contains
three terms,
\bq
S = S_V + S_{\overline{V}} + S_{\partial V},
\eq
where

\bq
S_V =  \int_V d^2x \overline{\psi} i\gamma_\mu\partial^\mu \psi + \ldots ,
\eq
\bq
S_{\overline{V}}=\int_ {\overline{V}} d^2x \frac{1}{2} (\partial_\mu \varphi)^2 + 
\ldots .
\eq
Here the dots, represent other possible contributions which we may add for
phenomenological reasons satisfying the above
restrictions. For the time being we will limit our discussion to the free
theory. The boundary condition is essential in making the connection between the
two regions. In $1+1$ dimensions we are guided by the bosonization rules, which
lead to 
\bq
S_{\partial V} = \int_{\partial V} d\Sigma_\mu \frac{1}{2} n^\mu 
 \overline{\psi}
e^{i\frac{\varphi\gamma _5}{f}}\psi ,
\eq
where the $\varphi$ decay constant $f=\frac{1}{4\pi}$, $d\Sigma _\mu$ is 
 an area
element and $n_\mu$ ($n^2 =-1$) the normal vector.

At the classical level the fermion satisfies
\bq
i\gamma^\mu \partial_\mu \psi = 0 ,
\eq
inside, while outside the boson obeys
\bq
\partial_\mu\partial^\mu \varphi = 0 ,
\eq
subject to the boundary conditions
\bq
i n_\mu \gamma^\mu \psi = -e^{i\frac{\gamma_5\varphi}{f}}\psi ,
\label{color}
\eq
\bq
n_\mu \partial^\mu \varphi =\frac{1}{2f} {\overline{\psi}} n_\mu \gamma^\mu
\gamma_5\psi .
\eq

From the boundary condition, Eq.\ref{color}, one may prove
\bq
n_\mu {\overline{\psi}} \gamma^\mu \psi = 0 ,
\eq
at the surface, which states that no flux of isoscalar current flows through 
the surface. But we are suppose to be describing at this stage a free fermion
theory! The crucial observation that leads to the solution of this apparent 
contradiction is that this classical result is invalidated by quantum mechanical
effects. The boundary condition generates a  quantum anomaly in the
fermion number current, i.e.,
\bq
dt \dot{Q} \sim \dot{\varphi} \sim d\Sigma \overline{\psi} n_\mu \gamma^\mu
\psi|_{\partial V} \sim t^\mu \partial_\mu ,
\eq
where $t_\mu$ is the tangential unit vector to the surface. In this way
we recover the full bosonization equation,
\bq
\partial_\mu \varphi = \frac{1}{2f}\overline{\psi} \gamma _\mu\gamma _5 \psi .
\eq
In two dimensions it turns out that
\bq
\varepsilon ^{\mu\nu}\partial_\nu \varphi = \overline{\psi} \gamma^\mu \psi
\eq
and this current is a topological current (trivially
conserved), whose charges define the different solitonic sectors. 
In $\overline{V}$ the fermions are described as solitons. The bag wall is
transparent for the fermions.

\subsection{Fermion charge leakage}

The interaction of the pseudoscalar field $\varphi$ with the fermion field $\psi$
gives rise to an anomaly and therefore the fermion number is not conserved
inside the bag. Physically what happens is that the amount of fermion number
$dQ$ corresponding to $\dot{\varphi}$ is pushed into the Dirac sea at the bag
boundary and so is lost from inside. This accumulated charge must be carried by
something residing in the meson sector. The meson field can only carry fermion
charge if it supports a soliton. In the present model we find that the fermion
charge $Q=1$ is partitioned as \cite{vento}
\bq
Q= 1 = Q_V + Q_{\overline{V}} ,
\eq
where
\bq
Q_V = 1 -\frac{\Theta}{\pi} ,
\eq
\bq
Q_{\overline{V}} = \frac{\Theta}{\pi} ,
\eq
with the chiral angle defined by
\bq
\Theta=\frac{\varphi}{f}|_{\partial V} 
\eq
We thus learn that the quark charge is partitioned into the bag and outside the
bag, without however any dependence of the total on the size or location of the
bag boundary. In the $1+1$-dimensional case, one can calculate other physical
quantities such as the energy, response functions, partition functions, and show
that the physics does not depend upon the presence of the bag. The complete
independence of the physics on the bag-size or -location is the realization of 
the Cheshire Cat Principle, and in two dimensions it holds exactly because of
the exact bosonization rule.

\subsection{Color anomaly}
\indent\indent Let us imitate the real world by incorporating a {\it color} 
charge. In the present case it will be sufficient with a $U(1)$ charge,
since the corresponding model, i.e., the Schwinger model, is confining . When
the anomaly at the boundary forces the quark to drawn into the Dirac sea, the
{\it color} charge of the quark vanishes too,
\bq
\dot{Q}_c = \frac{e}{2\pi} \frac{\dot{\varphi}}{f} .
\eq
The charge becomes non-conserved and hence gauge invariance is broken. To avoid
it, in a confining scenario (no gluons outside), we must introduce a
compensating charge at the surface. Thus the surface action changes to
\bq
S_{\partial V} = \int_{\partial V} d\Sigma \;[\frac{1}{2} \overline{\psi}
e^{i\gamma_5\varphi}{f} \psi -
\frac{e}{2\pi}\varepsilon^{\mu\nu}n_\mu A_\nu \frac{\varphi}{f}] .
\eq
The boundary conditions change in order to respect the gauge symmetry. This
consistency condition is crucial if the $CCP$ is to be respected.

Let us conclude this section with some comments which capture the full scope of
the $CCP$. The Schwinger model has a Coulomb confinement. The $CCP$ tells us,
that the bag does not confine! In the large bag limit, confinement is
described by means of a linearly rising Coulomb potential between the fermions.
In the small size limit, the boson acquires a mass
\bq
m^2=\frac{e^2}{\pi}.
\eq
It is this mass which confines them. In the bosonized Schwinger model the bosons
represent the fractionally charged color fermions.

\subsection{Approximate CCP in $3+1$-dimensions}

There is no exact $CCP$ in $3+1$-dimensions since fermion theories cannot be
bosonized exactly. The strategy will thus change. We will implement effective
equivalent bosonic theories by means  of Weinberg's theorem, i.e., symmetry
requirements and field theory principles, and will invoke the CCP as a
consistency condition of the bosonized theory.

We consider quarks confined within a volume $V$ which we take spherical for
definiteness, surrounded by a triplet of Goldstone bosons $\pi = \vec{\pi}\cdot
\frac{\vec{\tau}}{2}$ and the singlet
$\eta^\prime$ populating the outside volume $\overline{V}$. In such a scenario
the
action is given by the three terms \footnote{We shall just consider u and d
flavors. If strangeness is included the above lagrangian has to be supplemented
by the Wess-Zumino-Witten term.}
\bq
S = S_V +S_{\overline{V}} +S_{\partial V} ,
\eq
where
\bq
S_V = S_{QCD} ,
\eq
\bq
S_{\overline{V}} = S_{effective} =\frac{f^2}{4} \int_{\overline{V}}d^4x \;[
Tr(\partial _\mu
U^+)(\partial^\mu U) +\frac{m_\eta^2}{4N_f}\;Tr(\ln U -\ln U^+)^2] .
\eq
Here  $U$ is given by
$$
U=e^{\frac{i\eta}{f_0}} e^{\frac{2i\pi}{f}}
$$
and $f_0 = \sqrt{\frac{N_f}{2}} f$. We use $\eta$ to symbolize the
$\eta^\prime$ field.

The surface action is non trivial because it must contain the terms necessary 
to cancel the color anomaly \cite{NRWZ1} induced by the $\eta^\prime$ 
coupling. Its functional form is given by
\bq
\frac{1}{2} \int_{\partial V} d\Sigma ^\mu [n_\mu \overline{\psi} U_5\psi +
i\frac{g^2}{16\pi^2} K_\mu^5 \;Tr (\ln U^+ -\ln U)] ,
\label{cs}
\eq
where 
$$
U_5 =e^{i\frac{\eta \gamma_5}{f_0}} e^{\frac{2i\pi\gamma_5}{f}} ,
$$
and $K^5_\mu$ is the so called Chern-Simons current
\bq
K^5_\mu = \varepsilon^{\mu\nu\alpha\beta}\left(A^a_\nu F^a_{\alpha \beta}
-\frac{2}{3} gf^{abc} A_\nu^aA^b_{\alpha}A^c_{\beta}\right) .
\eq

The $CCP$ has been observed at the level of topological quantities, i.e., baryon
charge fractionation \cite{gjaffe} and approximately at the level of non
topological observables, i.e., masses, magnetic moments, etc \ldots
\cite{jackson,PV}. The explicit manifestation of the $CCP$ in the latter is
through some type of minimum sensitivity principle in terms of the bag radius.
Moreover the mean value about which observable are not sensitive to the radius
corresponds to the confinement scale $R\sim\frac{1}{\Lambda_{QCD}}$. 

\section{Anomaly and proton spin}

\indent\indent The anomalous suppresion of the first moment, $\Gamma^p_1$, of
the polarised
proton structure function $g_1^p$ has been the focus of intense theoretical
and experimental activity for nearly a decade. While it is now generally
accepted that the key to understanding this effect is the existence of the
chiral $U(1)$ anomaly in the flavor singlet pseudovector channel there are
several explanations reflecting different theoretical approaches to proton
structure. In here we analyze the phenomenon from the point of view of the
Cheshire Cat Principle. 

The starting point is the sum rule for the first moment, i.e.,
$$
\Gamma^P_1(Q^2) \equiv \int_0^1dx g_1^p(x,Q^2) =
\frac{1}{12}C_1^NS(\alpha_s(Q^2))\left(a^3 
+ \frac{1}{3} a^8\right) $$
\bq
+\frac{1}{9}C_1^S(\alpha_S(Q^2))a^0(Q^2) .
\eq
Here $a^3$, $a^8$ and $a^0(Q^2)$ are the form factors in the forward proton
matrix elements of the renormalised axial current, i.e.,
$$
<p,s|A^3_\mu|p,s> =s_\mu \frac{1}{2} a^3, \;\;\;\;\; <p,s|A^8_\mu|p,s> =s_\mu
\frac{1}{2\sqrt{3}} a^8, $$ 
and 
$$ 
<p,s|A^0_\mu|p,s> =s_\mu  a^0 ,
$$
where $p_\mu$ and $s_\mu$ are the momentum and the polarisation vector of the
proton. $a^3$ and $a^8$ can be chosen $Q^2$ independent and may be determined 
from the $\frac{G_A}{G_B}$ and $\frac{F}{D}$ ratios. $a^0(Q^2)$ evolves due to
the anomaly and its evolution can be decribed in the AB scheme \cite{ball} 
by
\bq
a^0(Q^2) = \Delta \Sigma - N_F \frac{\alpha_S(Q^2)}{2\pi} \Delta g(Q^2) .
\eq
Naive models or the $OZI$ approximation to $QCD$ lead at low energies to
\bq
a^0 \approx a^8 \approx 0.69 \pm 0.06 .
\eq
Experimentally\cite{forte}
\bq
a^0(\infty) = 0.10 \;^{+0.17}_{-0.10} .
\eq
The explanation of this unexpected behavior ranges from those authors 
attributing the whole effect to hadron structure
\cite{ellis}, to those attributing it to evolution \cite{shore,altarelli}. 
We shall discuss the problem from the point of view of hadron structure and
analyze if the Cheshire Cat Principle is realized within the formalism described
in
the previous section.

\subsection{Chiral bag formulation}
We next show that taking a proper account of the surface term, cf. Eq.
\ref{cs}, allows us to formulate a fully consistent
gauge invariant treatment of the flavor-singlet axial current (FSAC) 
matrix element \cite{RV}, which is related to the above observables. 
\bq
a^0(Q^2) = \frac{N_F}{M} <p,s|{\cal O}|p,s> ,
\eq
where
\bq
{\cal O} = \frac{\alpha_S}{8\pi} Tr(\tilde{G}_{\mu\nu} G^{\mu\nu}) +
\frac{1}{2N_F}(1-Z)\partial^\mu A^0_{\mu , bare} .
\eq
$Z$ is the renormalization constant of the current.

The equations of motion for the gluon and quark fields
inside and the $\eta^\prime$ field outside are the same as in
\cite{PVRB,PV}. However the boundary conditions on the surface now 
read
\bq
\hat{n}\cdot \vec{E}^a=-\frac{N_F g^2}{8\pi^2 f} \hat{n}\cdot \vec{B}^a 
\eta\label{E}
\eq
\bq
\hat{n}\times \vec{B}^a=\frac{N_F g^2}{8\pi^2 f} \hat{n}\times \vec{E}^a 
\eta\label{B}
\eq
and
\bq
\frac 12 \hat{n}\cdot({\overline{\psi}}{\vec{\gamma}}\gamma_5\psi)
=f \hat{n}\cdot{\vec{\d}} \eta +C \hat{n}\cdot \vec{K}\label{bc}
\eq
where $C=\frac{N_F g^2}{16\pi^2 }$ and $\vec{E}^a$ and $\vec{B}^a$ are,
respectively, the color electric and color magnetic fields. 
Here $\psi$ is the QCD quark field.

As it stands, the boundary condition for the $\eta^\prime$ field (\ref{bc})
looks gauge non-invariant because of the presence of the normal
component of the Chern-Simons current on the surface. However
this is not so. As shown in \cite{NRWZ1}, the term on the LHS of (\ref{bc})
is
not well-defined without regularization and when properly regularized,
say, by point-splitting, it can be written in terms of
a well-defined term which we will write as $\frac
12:{\overline{\psi}}\hat{n}\cdot
\gamma\gamma_5\psi:$ plus a gauge non-invariant term (see eq.(2) of 
\cite{NRWZ1}) which cancels exactly 
the second term on the RHS. The resulting boundary condition 
\bq
\frac 12\hat{n}\cdot:({\overline{\psi}}{\vec{\gamma}}\gamma_5\psi):
=f \hat{n}\cdot\vec{\d} \eta \label{bc2}
\eq
is then perfectly well-defined and gauge-invariant. However it is useless
as it stands since there is no simple way to evaluate the left-hand
without resorting to a model.
Our task in the chiral bag model is to 
express the well-defined operator $:(\overline{\psi}
\vec{\gamma}\gamma_5\psi):$ in terms of the bagged quark field $\Psi$. 
In doing this, our key strategy is to eliminate gauge-dependent surface terms
by the NRWZ surface counter term.

\subsection{Flavor-singlet axial current}
\indent\indent
Let us write the flavor-singlet axial current in the model as a sum of
two terms, one from the bag and the other from the outside populated by
the meson field $\eta^\prime$ (we will ignore the Goldstone pion fields for 
the moment)
\bq
A^\mu =A^\mu_B \Theta_B + A^\mu_M \Theta_M.\label{current}
\eq
We shall use the short-hand notations $\Theta_B=\theta (R-r)$ and
$\Theta_M=\theta (r-R)$ with $R$ being the radius of the bag which we shall
take to be spherical in this paper.
We demand that the $U_A (1)$ anomaly be given in this model by
\bq
\partial_\mu A^\mu = 
\frac{\alpha_s N_f}{2\pi}\sum_a \vec{E}^a \cdot \vec{B}^a \Theta_{B}+
f m_\eta^2 \eta \Theta_{M}.\label{ABJ}
\eq
Our task is to construct the FSAC in the chiral bag model
that is gauge-invariant and consistent with this anomaly equation. 
Our basic assumption is that in the nonperturbative sector outside of the
bag, the only relevant $U_A (1)$ degree of freedom is the massive $\eta^\prime$
field. (The possibility that there might
figure additional degrees of freedom in the exterior of the bag co-existing
with the $\eta^\prime$ and/or inside the bag 
co-existing with the quarks and gluons
will be discussed later.)
This assumption allows us to write
\bq
A^\mu_M = f\d^\mu \eta
\eq
with the divergence  
\be
\d_\mu A^\mu_M &=& fm_\eta^2 \eta.
\ee
Now the question is: what is the gauge-invariant and regularized
$A^\mu_B$ such that the anomaly (\ref{ABJ}) is satisfied?
To address this question, we rewrite the current (\ref{current})
absorbing the theta functions as
\bq
A^\mu=A_1^\mu +A_2^\mu
\eq
such that
\be
\partial_\mu A_1^\mu &=& f m_\eta^2 \eta \Theta_{M},\label{Dbag}\\
\partial_\mu A_2^\mu &=& 
\frac{\alpha_s N_f}{2\pi}\sum_a \vec{E}^a \cdot \vec{B}^a \Theta_{B}.
\label{Dmeson}
\ee
We shall deduce the appropriate currents in the lowest order in the 
gauge coupling constant $\alpha_s$
and in the cavity approximation for the quarks
inside the bag. 
\subsubsection{The ``quark'' current $A_1^\mu$}
\indent\indent
Let the bagged quark field be denoted $\Psi$. Then to the {\it lowest order}
in the gauge coupling and ignoring possible additional degrees of
freedom alluded above, the boundary condition (\ref{bc2}) is 
\bq
\frac 12\hat{n}\cdot({\overline{\Psi}}{\vec{ \gamma}}\gamma_5\Psi)
=f \hat{n}\cdot \vec{\d} \eta \label{bc3}
\eq
and the corresponding current satisfying (\ref{Dbag}) is
\bq
A_1^\mu=A_{1q}^\mu +A_{1\eta}^\mu
\eq
with
\be
A_{1q}^\mu &=& (\overline{\Psi} \gamma^\mu \gamma_5\Psi)\Theta_B,\label{A1in}\\
A_{1\eta}^\mu &=& f\d^\mu \eta \Theta_M.\label{A1out}
\ee
We shall now proceed to obtain the explicit form of
the bagged axial current operator.
In momentum space, the quark contribution is
\be
A^j_{1q}(q) &=& \frac{1}{2} \int d^3r e^{i \vec{q} \cdot \vec{r}}
\la N_{Bag}|\Psi^\dagger\sigma^j\Psi|N_{Bag}\ra\nonumber\\
&=& \left(a(q) \delta_{j\,k} +b(q)(3 \hat{q}_{j} \hat{q}_{k}
- \delta_{j\,k})\right) \la \frac{1}{2} \sum_{quarks} \sigma^k\ra\label{A1q}
\ee
where
\be
a(q)& =& N^2 \int dr r^2 (j_0^2(\omega r) -\frac{1}{3}j_1^2(\omega r))
j_0(q r),\label{MITga}\\
b(q) &=& \frac{2}{3} N^2 \int dr r^2 j_1^2(\omega r) j_2(q r)
\ee
where $N$ is the normalization constant of the (bagged) quark wave function.
In the limit that $q\rightarrow 0$ which is what 
we want to take for the axial charge, both terms are non-singular 
and only the $a(0)$ term survives, giving
\bq
A^j_{1q}(0) = g_{A,quark}^0  \la \frac{1}{2} \sum_{quarks} \sigma^j\ra
\label{1q}
\eq
where $g_{A,quark}^0$ is the singlet axial charge of the bagged quark
which can be extracted from (\ref{MITga}). In the numerical estimate
made below, we shall include the Casimir effects associated with
the hedgehog pion configuration to which the quarks are coupled
\cite{casimir,falomir}, so
the result will differ from the naive formula (\ref{MITga}).

To obtain the $\eta^\prime$ contribution, we take the $\eta^\prime$ 
field valid for 
a static source
\be
\eta (\vec{r}) = - \frac{g}{4\pi M} \int d^3r'\chi^{\dagger} \vec{S}
\chi \cdot \vec{\nabla} \frac{e^{-m_{\eta}|\vec{r} - \vec{r'}|}}
{|\vec{r} - \vec{r'}|}
\ee
where $g$ is the short-hand for the $\eta^\prime NN$ coupling constant,
$M$ is the nucleon mass, $\chi$ the Pauli spinor for the nucleon and 
$S$ the spin operator. The contribution to the FSAC is
\be
A^j_{1\eta}(q) &=& \int_{V_M} d^3r  e^{i \vec{q} \cdot \vec{r}}f 
\partial_j \eta,\nonumber\\
&=& (c(q) \delta_{j\,k} +d(q)(3 \hat{q}_{j} \hat{q}_{k}
- \delta_{j\,k})) \la \frac{1}{2} \sum_{quarks} \sigma^k\ra
\ee
with
\be
c(q) &=& \frac{fg}{2M}\int_R^\infty drr^2 \frac{e^{-m_\eta r}}{r} m_\eta^2 
j_0 (qr),\label{cq}\\
d(q) &=& -\frac{fg}{2M}\int_R^\infty dr \frac{e^{-m_\eta r}}{r}
[r^2 m_\eta^2 +3 (m_\eta r +1)] j_2 (qr).\label{dq}
\ee
In the zero momentum transfer limit\footnote{With however $m_\eta\neq 0$.
The limiting processes $q\rightarrow 0$ and $m_\eta\rightarrow 0$ do not
commute as we will see shortly.}, we have
\bq
A^j_{1\eta}(0) = \frac{g f}{2M}\left[(y_{\eta}^2 +2(y_{\eta}+1))\delta_{j\,k} -
y_{\eta}^2 \hat{q}_{j} \hat{q}_{k} \right] e^{-y_{\eta}} \la S^k\ra
\label{1eta}
\eq
where $y_\eta=m_\eta R$.

The boundary condition (\ref{bc3}) provides the relation between the 
quark and $\eta^\prime$ contributions. In the integrated form, (\ref{bc3})
is 
\bq
\int d\Sigma f x_3  \hat{r} \cdot\vec{\nabla} \eta = \int_{V_B} d^3 r
\frac{1}{2}  {\overline \Psi} \gamma_3 \gamma_5 \Psi
\eq
from which follows
\bq
\frac{g f}{M} = 3\, \frac{e^{y_\eta}}{y_\eta^2 + 2(y_\eta +1)}\, g_{A,quark}^0.
\label{ga2}
\eq
This is a Goldberger-Treiman-like formula relating the asymptotic pseudoscalar
coupling to the quark singlet axial charge.
From (\ref{1q}), (\ref{1eta}) and (\ref{ga2}), we obtain
\be
A_1^j=g_{A_1}^0 \la S^j\ra
\ee
with
\bq
g^0_{A_1}\label{A1}=\frac{gf}{3M}\frac{y_\eta^2 + 2(y_\eta +1)}{e^{y_\eta}}
=\frac 32 g_{A,quark}^0.
\label{ga3}
\eq
This is completely
analogous to the isovector axial charge $g_A^3$ coming from the bagged
quarks inside the bag plus the perturbative pion fields outside the bag.
Note that the singlet charge $g^0_{A_1}$ goes to zero when the bag is
shrunk to zero, implying that the coupling constant $g$ goes to
zero as $R\rightarrow 0$ as one can see from eq.(\ref{ga2}). 
This is in contrast to $g_A^3$ where
the axial charge from the bag ``leaks'' into the hedgehog pion
outside the bag and hence even when the bag shrinks to zero, the isovector
axial charge remains more or less constant in agreement with the 
Cheshire Cat\cite{hosaka}.

An interesting check of our calculation of $\vec{A}_1$
can be made by looking at the $m_\eta\rightarrow 0$ limit. From (\ref{A1q})
and (\ref{1eta}), we find that our current satisfies
\bq
\hat{q}\cdot \vec{A}_1 (0)=\frac{gf}{M}(y_\eta +1) e^{-y_\eta}\la \hat{q}\cdot
\vec{S}\ra\label{conserve}
\eq
which corresponds to eq.(\ref{Dbag}). Now eq.(\ref{Dbag}) is an operator
equation so one can take the limit $m_\eta\rightarrow 0$ and expect the
right-hand side to vanish, obtaining $\hat{q}\cdot\vec{A}_1\rightarrow 0$.
Equation (\ref{conserve}) fails to satisfy this. The reason for this
failure is that the $q\rightarrow 0$ and $m_\eta\rightarrow 0$ limits do not
commute. To obtain the massless limit, one should take the $\eta^\prime$ mass
to go to zero first.

Before taking the zero-momentum limit,  the expression for $c(q)$ for
the $\eta$ field, (\ref{cq}), is
\bq
c(q) = \frac{fg}{3M}\left( \frac{m_\eta^2}{q^2}
\frac{e^{-y_\eta}(\cos({qr}) +\frac{m_\eta}{q}\sin({qr}))}{1 +
\frac{m_\eta^2}{q^2}}\right)
\eq
which vanishes in the  $m_\eta\rightarrow 0$ limit. On the other hand,
the $d (q)$, (\ref{dq}),  which before taking the zero-momentum limit,
is of the form 
\be
d(q) &=& -\frac{fg}{2M}(\frac{e^{-y_\eta}(y_\eta^2 + 3y_\eta +3)}
{qR} j_1(qR)\nonumber\\
&&- \frac{m_\eta^2}{q^2}e^{-y_\eta}(y_\eta +1)j_0(qR) +
\frac{m_\eta^4}{q^4}
\frac{e^{-y_\eta}(\cos({qr}) +\frac{m_\eta}{q}\sin({qr}))}{1 +
\frac{m_\eta^2}{q^2}})
\ee  
becomes in the $m_\eta\rightarrow 0$ limit
\bq
-\frac{fg}{2M}\frac{j_1(qR)}{qR}.
\eq
Adding the  quark current (\ref{A1q}) in the $q\rightarrow 0$ limit, we get
\bq
A_1^j(0) = \frac{fg}{M}(\delta^{jk} -\hat{q}^j\hat{q}^k) S_k
\eq
which satisfies the conservation relation. This shows that our formulas
are correct.

Note that the massless limit leads to
\bq
g^0_A = \frac{gf}{2M}
\eq
which is the $U(1)$ G-T relation of Shore and Veneziano in the $OZI$ limit
\cite{shore}. In this formula one can recover the target independent factor
which is determined by $f \sim \sqrt{\chi^\prime (0)}$ \footnote{Here $\chi$
represents the topological susceptibility of the $QCD$ vacuum.},while $g
\sim\Gamma_{\varphi_5 NN}$ carries the target dependent contribution.

\subsubsection{The gluon current  $A_2^\mu$}
\indent\indent
The current $A_2^\mu$ involving the color gauge field is very intricate
because it is not possible in general
to write a gauge-invariant dimension-3 local operator
corresponding to the singlet channel. We will see however that it is
possible to obtain a consistent {\it axial charge} within the model. Here
we shall
calculate it to the lowest nontrivial order in the gauge coupling constant.
In this limit, the right-hand sides of the boundary conditions (\ref{E})
and (\ref{B}) can be dropped, reducing to the original MIT boundary 
conditions \cite{mit}. Furthermore the gauge field decouples from the other
degrees of
freedom precisely because of the color anomaly condition that prevents
the color leakage, namely, the condition (\ref{bc2}). In its absence, this
decoupling could not take place in a consistent way.\footnote{To higher
order in the gauge coupling, the situation would be a lot more complicated.
A full Casimir calculation will be required to assure the consistency of
the procedure. This problem will be addressed in a future publication.}

We start with the divergence relation
\bq
\partial_\mu A^\mu_2 =
\frac{\alpha_s N_f}{2\pi}\sum_a \vec{E}^a \cdot \vec{B}^a \Theta_{V_B}.
\label{an2}
\eq
In the lowest-mode approximation, the color electric and magnetic fields are
given by
\bq
\vec{E}^a = g_s \frac{\lambda^a}{4\pi} \frac{\hat{r}}{r^2} \rho (r)
\label{ef}
\eq
\bq
\vec{B}^a = g_s \frac{\lambda^a}{4\pi}\left( \frac{\mu (r)}{r^3}(3 \hat{r}
\vec{\sigma} \cdot \hat{r} - \vec{\sigma}) + (\frac{\mu (R)}{R^3} + 2 M(r))
\vec{\sigma}\right)
\label{bf}
\eq
where $\rho$ is related to the quark scalar density $\rho^\prime$ as
\bq
\rho (r)=\int_\Gamma^r ds \rho^\prime (s)\label{density}\nonumber
\eq
and $\mu, M$ to the vector current density
\be
\mu (r) &=& \int_0^r ds \mu^\prime (s),\nonumber\\ 
M (r)&=& \int_r^R ds \frac{\mu^\prime (s)}{s^3}.\nonumber
\ee
The lower limit $\Gamma$ usually taken to be zero in the MIT bag model
will be fixed later on. It will turn out that what one takes for $\Gamma$
has a qualitatively different consequence on the Cheshire-Cat property
of the singlet axial current.
Substituting these fields into the RHS of eq.(\ref{an2}) leads to
\bq
\vec{q} \cdot \vec{A_2} = \frac{8 \alpha _s^2 N_f}{3\pi} \vec{\sigma} \cdot
\hat{q} \int_0^R dr \rho (r)\left(2\frac{\mu(r)}{r^3} + \frac{\mu
(R)}{R^3} + 2M(r)\right) j_1(qr)
\eq
where $\alpha_s = \frac{g^2_s}{4\pi}$ and we have used  $ \sum_{i\neq j}\sum_a
\lambda^a_i\lambda^a_j = - \frac{8}{3}$ for the baryons\footnote{Here we 
are making the usual assumption as in ref.\cite{jaffe} that the $i=j$ terms
in the color factor are to be excluded from the contribution on the ground
that most of them go into renormalizing the single-quark axial charge.
If one were to evaluate the color factor without excluding the 
diagonal terms using only the lowest mode, 
the anomaly term would vanish, which of course is incorrect. 
As emphasized in \cite{jaffe}, there may be residual finite
contribution with $i=j$ but no one knows how to compute this and so
we shall ignore it here. It may have to be carefully considered in a
full Casimir calculation yet to be worked out.}.

In order to calculate the axial charge, we  take the zero momentum limit and
obtain
\bq
\lim_{q \rightarrow 0} \vec{A_2}(\vec{q}) = \frac{8 \alpha_s^2 N_f}{9\pi}
\tilde{A}_2(R)\vec{S}
\label{A2}
\eq
where
\bq
\tilde{A}_2(R) = \int_0^R r dr \rho (r)\left(2M(r) + \frac{\mu(R)}{R^3} +
2\frac{\mu(r)}{r^3}\right)\equiv 2 \int^R_0 dr r\rho(r) \alpha(r).\label{A}
\eq
The quantity $\alpha(r)$ is defined for later purposes.
It is easy to convince oneself that (\ref{A2}) is gauge-invariant, i.e.,
it is $\propto \int_{V_B} d^3r \vec{r} \sum_a \vec{E}^a \cdot \vec{B}^a$
which is manifestly gauge-invariant. The result (\ref{A2}) was previously
obtained in \cite{hatsudazahed}.

\subsubsection{The two-component formula}
\indent\indent
The main result of this paper can be summarized in terms of 
the two component-formula for the singlet axial charge (with $c_1=0$),
\bq
g_A^0=g_{A_1}^0 +g_{A_2}^0 =\frac{3}{2} g^0_{A,quarks} + 
\frac{8 \alpha_s^2 N_f}{9\pi} A_2(R).\label{SV}
\eq
The first term is the ``matter'' contribution (\ref{A1})
and the second the gauge-field contribution (\ref{A2}). 
This is the chiral-bag version of Shore-Veneziano 
formula\cite{shore,schechter} relating
the  singlet axial charge to a sum of an $\eta^\prime$ 
contribution and a glueball
contribution,
\bq
g_A^0 = \frac{fg_{\eta NN}}{2M} + \frac{f^2m_\eta^2}{2N_f} g_{GNN}(0)
\eq
It is immediate to realize that there exists a one to one correspondence 
between the two expressions. The first term establishes a microscopic
description of the $\eta^\prime$ coupling in a quarkish scenario \cite{RV}. The
second term reformulates the glueball contribution in terms of the axial anomaly
inside the bag.

\section{Results}
\indent\indent
In this section, we shall make a numerical estimate of (\ref{A1}) and 
(\ref{A2}) in the approximation that is detailed above. In evaluating
(\ref{A1}), we shall take into account the Casimir effects due to the
hedgehog pions but ignore the effect of the $\eta^\prime$ field on
the quark spectrum. The interaction between the internal and external 
degrees of freedom occurs at the surface.  Our approximation consists 
of  neglecting in the expansion of the boundary condition  in powers of 
$\frac{1}{f}$ all $\eta$ dependence, i.e.
\bq
i\hat{r}\cdot \vec{\gamma}\Psi =
e^{i\gamma_5 \vec{\tau}\cdot \hat{r}\frac{\varphi(\vec{r})}{f_\pi}}
e^{i\gamma_5 \frac{\eta}{f}}\Psi \sim 
e^{i\gamma_5 \vec{\tau}\cdot \hat{r}\frac{\varphi(\vec{r})}{f_\pi}}
\Psi
\label{qbc}
\eq 
This approximation is justified by the massiveness of the $\eta^\prime$
field in comparison to the Goldstone pion field that supports the
hedgehog configuration, $\varphi$. Within this approximation, 
we can simply take the numerical results from \cite{PVRB,PV} 
changing only the overall constants in front.

The same is true with the gluon contribution. To the lowest order
in $\alpha_s$, the equation of motion  for the gluon field is the
same as in the MIT bag model. This is easy to see, since the modified 
boundary conditions eqs.(\ref{E}) and (\ref{B}) become
\bq
\hat{r}_i G^{i \mu} = - \frac{\alpha_s N_F}{2 \pi}\frac{\eta}{f}\hat{r}_i 
{\tilde G}^{ i \mu} \sim 0.
\eq
The only difference from the MIT model is that
here the quark sources for the gluons are modified by the hedgehog pion
field in (\ref{qbc}). Again the results can be taken from 
\cite{PVRB,PV} modulo an 
overall numerical factor. 

In evaluating the anomaly contribution (\ref{A}), we face the same problem
with the monopole component of the ${{\vec E}^a}$ field as in 
\cite{PVRB,PV}. If we write 
\bq
{\vec{E}}_i^a (r)=f(r) \hat{r} \lambda_i^a
\eq
where the subscript $i$ labels the $i$th quark and $a$ the color,
the $f(r)$ satisfying the Maxwell equation is
\bq
f(r)=\frac{1}{4\pi r^2}\int^r_\Gamma ds \rho^\prime (s)
\equiv \frac{1}{4\pi r^2}\rho(r).\label{fr}
\eq
If one takes only the valence quark orbit -- which is our approximation,
then $\rho^\prime$ in the chiral bag takes the same form as in the
MIT model. However the quark orbit is basically modified by the
hedgehog boundary condition, so the result is of course not the same.
The well-known difficulty here is that the bag boundary condition for
the monopole component
\bq
\hat{ r }\cdot {{\vec{E}}_i^a}=0,\ \ \ \ {\rm at}\ \ r=R
\eq
is not satisfied for $\Gamma\neq R$. Thus as in \cite{PVRB,PV},
we shall consider both $\Gamma=0$ and $\Gamma=R$.\footnote{These choices
describe two opposite scenarios for confinement.
The standard MIT solution\cite{mit} proposes a mechanism 
for satisfying the boundary condition of the electric field based on the 
color matrices which does not impose any restriction on $\Gamma$. The value
$\Gamma=0$ was chosen because the field becomes the typical field of a charged 
sphere in the abelianized theory. The spatial structure of the 
electric field is locally non-confining. The procedure introduces an asymmetry 
in the way confinement is realized between the electric and magnetic fields.
This asymmetry leads to a peculiar treatment of the self-energy
terms in the model.
The $\Gamma=R$ solution \cite{PVRB} describes the opposite scenario.  Color
electric screening occurs explicitly. As one moves away from the center the
color charge decreases so that at the surface the color electric charge 
of the bag is zero. Abelianizing the theory, i.e., eliminating the color
matrices, the mechanism can be visualized as a pointlike charge (-Q) at
the origin superposed to the conventional spherical charge distribution,
whose total charge is +Q. In this way one sees explicitly that the field 
lines for the quarks are moving towards the interior of the sphere,
contrary to what happens in the other scenario. Now the deeper question
arises: what is the microscopic mechanism producing this scenario? 
In our opinion this is a spherical modelization of the flux tube.}

The existence of a solution which satisfies explicitly and
locally the boundary condition suggests an approach different from
the one in the original MIT calculation \cite{mit}, 
where the boundary condition of the electric field was imposed as an 
expectation value with respect to the physical hadron state. 
In \cite{mit}, the $E$ and $B$ field contributions 
to the spectrum were treated on a completely different footing. While in the 
former the contribution arising from the quark self-energies was included, 
thereby leading to the vanishing of the color electric energy, in the
latter they were not. This gave the color magnetic energy for the source of the 
nucleon-$\Delta$ splitting. We have performed a calculation for the energy with
the explicitly confined $E$ and $B$ field treated in a symmetric fashion
\cite{PV}. Although in this calculation the contribution of the color-electric 
energy was non-vanishing, it was found not to affect the nucleon-$\Delta$ mass 
splitting, and therefore could be absorbed into a small change of the 
unknown parameters, i.e., zero point energy, bag radius, bag pressure etc.
As we shall see shortly, the two ways of treating the confinement
with $\Gamma=R$ and $\Gamma=0$ give qualitatively different results
for the role of the anomaly. One could consider therefore that the
singlet axial charge offers a possibility of learning something about
confinement within the scheme of the chiral bag. At present, 
only in heavy quarkonia \cite{vz} does one have an additional
handle on these operators.

The numerical results for both cases are given in Table 1.

\begin{table}[tbh]
\caption{The flavor-singlet axial charge of the
proton as a function of radius $R$ and the chiral angle $\theta$. The
column labeled $g_{A_1}^0$ corresponds to the total contribution from
the quarks inside the bag and $\eta^\prime$ outside the bag (eq.(\ref{A1})) and
$g_{A_2}^0 (\Gamma=R)$  and ${g_{A_2}^0} (\Gamma=0)$ to the gluon contribution
eq.(\ref{A2}) evaluated with $\Gamma=R$ and
$\Gamma=0$ in (\ref{fr}), respectively.  The parameters are: $\alpha_s=2.2$, 
$m_\eta=958$ MeV and $f=93$ MeV. The row with $R=\infty$ corresponds to the
unrealistic (and extreme) case of
an MIT bag model with the same parameters for the same degrees of
freedom but containing {\it no pions}.}
\vspace{0.2cm}
\begin{center}
\footnotesize
\begin{tabular}{|c|c|c|c|c|c|c|}\hline
R(fm)&$\theta/\pi$&$g_{A_1}^0$&$g_{A_2}^0 (\Gamma=R)$&$g_{A_2}^0 (\Gamma=0)$&
$g_A^0 (\Gamma=R)$&$g_A^0 (\Gamma=0)$\\ \hline
0.2&-0.742&0.033&-0.015&0.009&0.018&0.042\\
0.4&-0.531&0.164&-0.087&0.046&0.077&0.210\\
0.6&-0.383&0.321&-0.236&0.123&0.085&0.444\\
0.8&-0.277&0.494&-0.434&0.232&0.060&0.726\\
1.0&-0.194&0.675&-0.635&0.352&0.040&1.027\\ \hline
$\infty$&0.00&0.962& -1.277&0.804 &-0.297 &1.784 \\ \hline
\end{tabular}
\end{center}
\end{table}

\section{Discussion}
\indent\indent
The quantity we have computed here is relevant to two physical issues:
the so-called ``proton spin" issue and the Cheshire-Cat
phenomenon in the baryon structure. A more accurate result
awaits a full Casimir calculation which appears to be non-trivial.
However we believe that the qualitative feature of the given model
with the specified degrees of freedom will not be significantly
modified by the full Casimir effects going beyond the lowest order
in $\alpha_s$.

In the current understanding of the polarized structure functions of the
nucleon, the FSAC matrix element or the flavor-singlet axial charge of
the proton is related to the singlet axial charge \cite{ellis,altarelli}
\bq
a_0(Q^2) = \Delta \Sigma (1,Q^2) - N_F\frac{\alpha_s(t)}{2\pi}\Delta g(1,Q^2)
\eq
The presently available analyses give \cite{altarelli}
\be
a_0(\infty) = 0.10\pm 0.05({\rm exp})\pm^{0.17}_{0.11}({\rm th})
=0.10 \pm^{0.17}_{0.11} \label{altarelli}
\ee
Evolution to lower momenta will increase this value according to a
multiplicative factor (see Fig. 5 in ref.\cite{forte}), which is not large
as long as one remains in the regime where the perturbative expansion is valid. 
Our predictions for $g_A^0$ -- which can be compared with $a_0(\mu_0^2)$, where
$\mu_0^2$ is the hadronic scale at which the model is defined \cite{jr} --
differ drastically depending upon whether one takes
$\Gamma=0$ for which the color electric monopole field satisfies {\it only
globally} 
the boundary condition at the leading order (that is, as a matrix element 
between color-singlet states) as in the standard 
MIT bag-model phenomenology or $\Gamma=R$ which makes
the boundary condition satisfied locally.  The former configuration severely
breaks the Cheshire Cat with the bag radius $R$ constrained to
less  than $0.5$ fm (``little bag scenario") to describe the 
empirical value (\ref{altarelli}). This is analogous to what Dreiner, 
Ellis and Flores
\cite{DEF} obtained. 

On the other hand, the configuration with $\Gamma=R$ which
we favor
leads to a remarkably stable Cheshire Cat in consistency with other
non-anomalous processes where the Cheshire Cat is seen to hold 
within, say, 30\% \cite{PV,hosaka}. The resulting singlet axial charge
$g_A^0 <0.1$ is entirely consistent with (\ref{altarelli}).
One cannot however take the near zero
value predicted here too literally since the value taken for $\alpha_s$
is perhaps too large.  Moreover other  short-distance degrees of
freedom not taken into account in the model (such as the light-quark vector 
mesons and other massive mesons)
can make a non-negligible additional contribution\cite{schechter}. What is 
noteworthy is that there is a large cancellation between
the ``matter" (quark and $\eta^\prime$) contribution and the gauge field 
(gluon) contribution in agreement with the interpretation
anchored on $U_A (1)$ anomaly\cite{altarelli}.

As mentioned above -- and also noted in \cite{PVRB,PV}, the electric monopole
configuration with $\Gamma=R$ is non-zero at the origin and hence is 
ill-defined there. This feature does not affect, however, other phenomenology
as shown in \cite{PV}. We do not know yet
if this ambiguity can be avoided if other multipoles
and higher-order and Casimir effects are included in a consistent way.
This caveat notwithstanding, it seems reasonable to conclude from the result 
that if one accepts that the
singlet axial charge is  small {\it because of the cancellation}
in the two-component formula and if in addition
one demands that the Cheshire Cat hold
in the $U_A (1)$ channel {\it as in other non-anomalous sectors}, we are led
to (1) adopt the singular monopole configuration that  satisfies
the boundary condition {\it locally} and (2) to the possibility
that within the range of
the bag radius that we are considering, the $\eta^\prime$ is primarily
quarkish.

The fact that the $CCP$ is realized in a specific dynamical way, requires
further investigations. It would seem natural that, if the effective theory
chosen to describe the low energy properties where close to reality, the $CCP$
would arise in a more natural fashion. Two have been our main simplifying
assumptions. In the exterior we have taken the minimal theory containing the
required degrees of freedom. In the interior we have performed an approximate
calculation which obviates the full content of the couping of gluons and
$\eta^\prime$. The $CCP$ will become a fundamental principle of hadron
structure if it survives in a natural way the complete calculation.

\section*{Acknowledgments}
The financial support of CPT and the warm hospitality of Profs. D. P. Min and
B.Y. Park are gratefully acknowledged. My gratitude goes also to Dr. F. Cano for
a careful reading of the manuscript. The friendly scientific atmosphere of this 
APCTP meeting made my attendance a most profitable experience. 
This research has been
supported by DGICYT-PB94-0080 and TMR programme of the European Commision 
ERB FMRX-CT96-008.

Almost twenty years ago I was introduced to Mannque by Gerry Brown. The
{\it Little Bag} was being developed and I joined in. I recall those days with
pleasure: I was young and the physics was great!.  Since then I have
collaborated with Mannque on some occasions, during which I was able to  
appreciate his fascination for
physics. I wish him well, in this very merited celebration, and hope that our 
joint ventures continue for a long time.

\end{document}